\documentclass[runningheads]{llncs}
\usepackage[framemethod=tikz]{mdframed}
\usepackage{lipsum}
\usepackage{graphicx}
\usepackage[many]{tcolorbox}
\usetikzlibrary{calc}

\usepackage{amsthm}
\usepackage{amsmath,amssymb, graphicx}
\usepackage[ruled,linesnumbered]{algorithm2e}
\usepackage{algorithmic}

\usepackage{paralist} 
\usepackage{color}
\usepackage{xcolor}
\usepackage{multirow}
\usepackage{rotating}

\usepackage[mathscr]{eucal}
\usepackage{amsbsy}
   
\usepackage{subfigure}
\usepackage{booktabs}

\usepackage{xcolor}
\usepackage{tikz}
\usetikzlibrary{snakes}
\usepackage{multirow}
\usepackage{color}
\usepackage{breqn}

\usepackage{epstopdf}
\usepackage{balance}
\usepackage{tablefootnote}
\usepackage{empheq} 
\usepackage{moresize}
\usepackage{tabularx}

\usepackage{enumitem}
\setlist{nolistsep}

\usepackage[font=small,labelfont=bf,skip=0pt]{caption}

\usepackage{url}

\renewcommand{\algorithmicrequire}{\textbf{Input:}}

\newcounter{ALC@tempcntr}

\renewcommand{\algorithmicrequire}{\textbf{Input:}}

\usepackage{xspace}

\newcommand{\svm}{\texttt{tf\underline{\hspace{.1in}}idf\slash SVM}\xspace}
\newcommand{\ourdata}{\texttt{Twitter}\xspace}
\newcommand{\politifact}{\texttt{Politifact}\xspace}
\newcommand{\TTA}{\texttt{TTA}\xspace}
\newcommand{\HTA}{\texttt{HTA}\xspace}
\newcommand{\Tags}{\texttt{TAGS}\xspace}
\newcommand{\CMTF}{\texttt{CMTF}\xspace}
\newcommand{\ACMTF}{\texttt{ACMTF}\xspace}
\newcommand{\CMTFpp}{\texttt{CMTF++}\xspace}

\newcommand{\CPJIVE}{\texttt{HiJoD}\xspace}
\newcommand{\CPSVD}{\texttt{CP-SVD}\xspace}
\newcommand{\CPICA}{\texttt{CP-ICA}\xspace}
\newcommand{\CP}{\texttt{CP}\xspace}
\newcommand{\JICA}{\texttt{JICA}\xspace}
\newcommand{\JIVE}{\texttt{JIVE}\xspace}
\newcommand{\SVD}{\texttt{SVD}\xspace}
\newcommand{\docvec}{\texttt{Doc2Vec\slash SVM}\xspace}
\newcommand{\fasttext}{\texttt{fastText}\xspace}
\newcommand{\tfidf}{\texttt{tf\underline{\hspace{.1in}}idf}\xspace}
\newcommand{\GloVeLSTM}{\texttt{GloVe\slash LSTM}\xspace}
\newcommand{\TTABP}{\texttt{TTA\slash BP}\xspace}

\newcommand{\norm}[1]{\left\lVert#1\right\rVert}

\usepackage{fancyvrb}

\begin{document}
\title{Semi-Supervised Multi-aspect Detection of Misinformation using \underline{Hi}erarchical \underline{Jo}int  \underline{D}ecomposition}%\thanks{Supported by organization x.}}
\author{
Sara Abdali\inst{1} \and\\ 
Neil Shah\inst{2} \and \\
Evangelos E. Papalexakis \inst{1}
 }
 \institute{Department of Computer Science and Engineering
University of California Riverside
900 University Avenue, Riverside, CA, USA\\ \email{sabda005@ucr.edu} \email{epapalex@cs.ucr.edu}
\and Snap Inc. \email{nshah@snap.com}}
\maketitle

\begin{abstract}
Distinguishing between misinformation and real information is one of the most challenging problems in today's interconnected world. The vast majority of the state-of-the-art in detecting misinformation is fully supervised, requiring a large number of high-quality human annotations. However, the availability of such annotations cannot be taken for granted, since it is very costly, time-consuming, and challenging to do so in a way that keeps up with the proliferation of misinformation.
In this work, we are interested in exploring scenarios where the number of annotations is limited. In such scenarios, we investigate how to tap into a diverse number of resources that characterize a news article, henceforth referred to as ``aspects'' can compensate for the lack of labels. In particular, our contributions in this paper are twofold: 1) We propose the use of three different aspects: article content, context of social sharing behaviors, and host website/domain features, and 2) We introduce a principled tensor based embedding framework that combines all those aspects effectively.
We propose \CPJIVE a 2-level decomposition pipeline which not only outperforms state-of-the-art methods with F1-scores of 74\% and 81\% on \ourdata and \politifact datasets respectively, but also is an order of magnitude faster than similar ensemble approaches.
\end{abstract}
\section{Introduction}
\label{sec:intro}

In recent years, we have experienced the proliferation of websites and outlets that publish and perpetuate misinformation. With the aid of social media platforms, such misinformation propagates wildly and reaches a large number of the population, and can, in fact, have real-world consequences. Thus, understanding and flagging misinformation on the web is an extremely important and timely problem, which is here to stay.
 
\par There have been significant advances in detecting misinformation from the article content, which can be largely divided into knowledge-based fact-checking and style-based approaches;  \cite{kumar2018false,shu2017fake} survey the landscape.
Regardless of the particular approach followed, the vast majority of the state-of-the-art is fully supervised, requiring a large number of high-quality human annotations in order to learn the association between content and whether an article is misinformative. For instance, in \cite{Castillo:2011:ICT:1963405.1963500}, a decision-tree based algorithm is used to assess the credibility of a tweet, based on Twitter features. In another work, Rubin et al. \cite{Rubin:2016} leverage linguistic features and a SVM-based classifier to find misleading information. Similarly, Horne and Adali \cite{Horne:2017} apply SVM classification on content-based features. There are several other works \cite{Gupta:Han:2012,NewsVerif,Jin:2014} for assessing credibility of news articles, all of which employ propagation models in a supervised manner.
\begin{figure}[t!]
\begin{center}
\includegraphics[width = 0.6\linewidth]{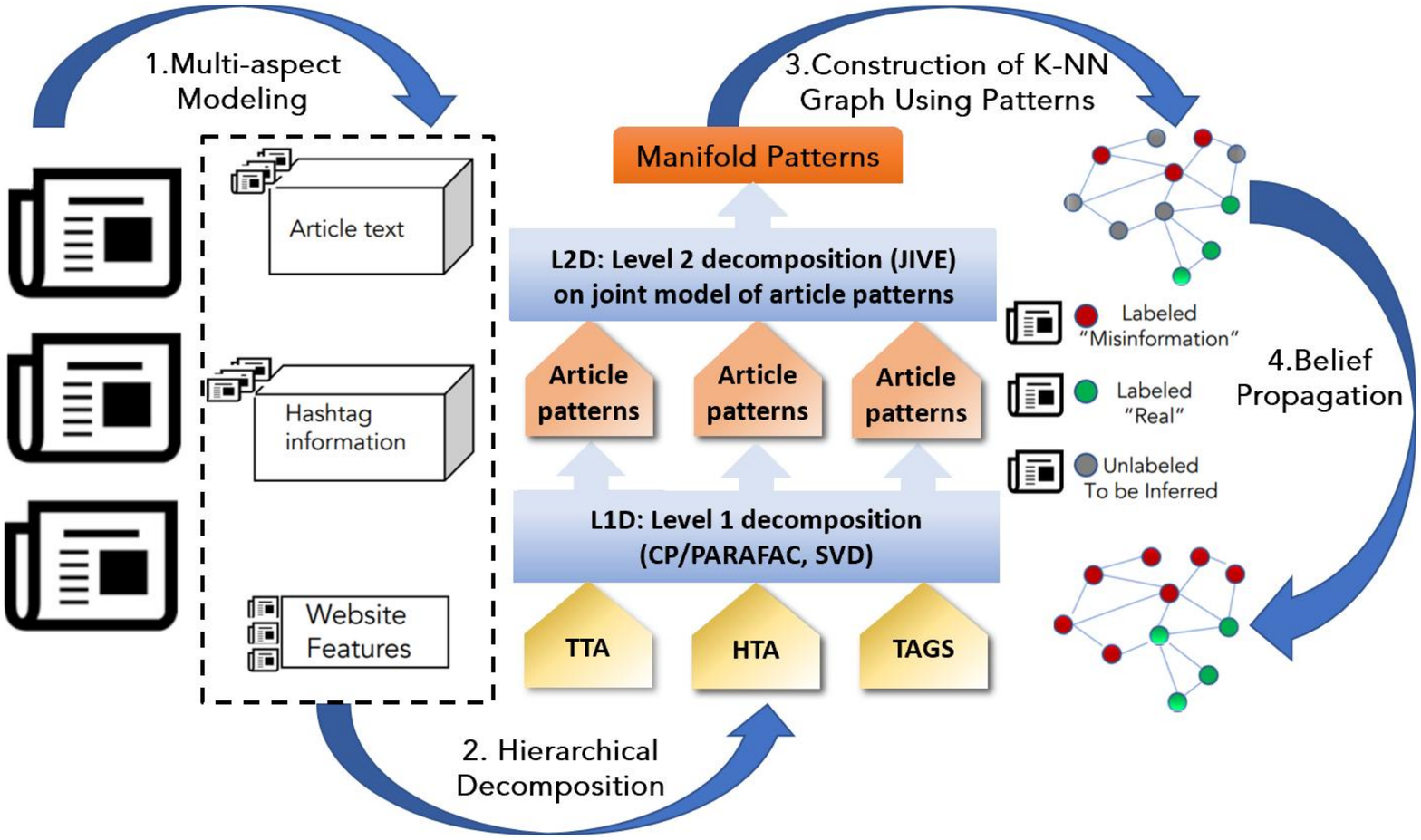}
\end{center}
        \caption{Overview: We propose using three aspects1: a) content, b) social context (in the form of hashtags), and c) features of the website serving the content. We, further, propose \CPJIVE hierarchical approach for finding latent patterns derived from those aspects, generate a graphical representation of all articles in the embedding space, and conduct semi-supervised label inference of unknown articles.}
        \label{fig:overview}
\end{figure}
Collecting human annotation for misinformation detection is a complicated and time consuming task, since it is challenging and costly to identify human experts who can label news articles devoid of their own subjective views and biases, and possess all required pieces of information to be a suitable ``oracle.'' However, there exist crowd-sourced schemes such as the browser extension ``BS Detector'' \footnote{http://bsdetector.tech/} which provide coarse labels by allowing users to flag certain articles as different types of misinformation, and subsequently flagging the entire source/domain as the majority label. Thus, we are interested in investigating methods that can compensate for the lack of large amounts of labels with leveraging different signals or aspects that pertain to an article. 
A motivating consideration is that we as humans empirically consider different aspects of a particular article in order to distinguish between misinformation and real information. For example, when we review a news article on a web site, and we have no prior knowledge about the legitimacy of its information, we not only take a close look at the content of the article but also we may consider how the web page looks (e.g. does it look ``professional''? Does it have many ads and pop-up windows that make it look untrustworthy?). We might conclude that untrustworthy resources tend to have more ``messy'' web sites in that they are full of ads, irrelevant images, pop-up windows, and scripts. Most prior work in the misinformation detection space considers only one or two aspects of information, namely content, headline or linguistic features. 
In this paper, we aim to fill that gap by proposing a comprehensive method that emulates this multi-aspect human approach for finding latent patterns corresponding to different classes, while at the same time leverages scarce supervision in order to turn those insights into actionable classifiers. 
In particular, our proposed method combines multiple aspects of article, including (a) article content (b) social sharing context and (c) source webpage context each of which is modeled as a tensor/matrix. The rationale behind
using tensor based model rather than state-of-the-art approaches like deep learning methods is that, such approaches are mostly supervised methods and require considerable amount of labeled data for training. On the contrary, we can leverage tensor based approaches to find meaningful patterns using less labels. Later on, we will compare the performance of tensor-based modeling against deep learning methods when there is scarcity of labels. We summarize the contributions as follows:
\begin{itemize}[noitemsep]
   \item {\bf Leveraging different aspects of misinformation}:
  In this work, we not only consider content-based information but also we propose to leverage multiple aspects for discriminating misinformative articles. In fact, we propose to create multiple models, each of which describes a distinct aspect of articles.   
    \item {\bf A novel hierarchical tensor based ensemble model}: We leverage a hierarchical tensor based ensemble method i.e., \CPJIVE to find manifold patterns that comprise multi-aspect information of the data.

	\item {\bf Evaluation on real data}: We extensively evaluate \CPJIVE on two real world datasets i.e., \ourdata and \politifact. \CPJIVE not only outperforms state-of-the-art alternatives with F1 score of 74\% and 81\% on above datasets respectively, but also is significantly faster than similar ensemble methods. We make our implementation publicly available\footnote{\scriptsize https://github.com/Saraabdali/HiJoD-ECMLPKDD} to promote the reproducibility.
	
\end{itemize}

\section{Problem Formulation}
Considering the following formulation of misinformation detection problem:
\tcbset{colback=gray!30!white,colframe=gray!70!white}
\begin{tcolorbox}[width=1\linewidth]
    \textbf{Given} $N$ articles with associated
    \iffalse
    \begin{compactitem}
    \item article text
    \item social sharing context (hashtags that are used when sharing the article)
    \item HTML source of the article's publisher webpage, and
    \item binary (misinformative/real) labels for $p\%$ of articles,
    \end{compactitem} 
    \fi
    1) article text,
    social sharing context (hashtags that are used when sharing the article)
    2) HTML source of the article's publisher webpage, and, 
    3) binary (misinformative/real) labels for $p\%$ of articles,
    \textbf{Classify} the remaining articles into the two classes.
\end{tcolorbox}

At a high level, we aim to demonstrate the predictive power of incorporating multiple aspects of article content and context on the misinformation detection task, and consider doing so in a low-label setting due to practical challenges in data labeling for this task.  Our proposed method especially focuses on (a) multi-aspect data modeling and representation choices for downstream tasks, (b) appropriate triage across multiple aspects, and (c) utility of a semi-supervised approach which is ideal in sparse label settings.  In the next section, we detail our intuition and choices for each of these components.

 We aim to develop a manifold approach which can discriminate misinformative and real news articles by leveraging content-based information in addition to other article aspects i.e. social context and publisher webpage information.  To this end, we propose using tensors and matrices to analyze these aspects jointly. We develop a three-stage approach: (a) 
\emph{multi-aspect article representation}: we first introduce three feature context representations (models) which describe articles from different points of view (aspects), (b) \emph{Manifold patterns finding using a hierarchical approach}:
In proposed \CPJIVE, we first decompose each model separately to find the latent patterns of the articles with respect to the corresponding aspect, then we use a strategy to find shared components of the individual patterns (c) \emph{semi-supervised article inference}: finally, we focus on the inference task by construing a K-NN graph over the resulted manifold patterns and propagating a limited set of labels.   %\reminder{V: More Fig. 2 here}

\section{Proposed Methodology}
\subsection{Multi-Aspect Article Representation}
We first model articles with respect to different aspects. We suggest following tensors/matrix to model content, social context and source aspects respectively:
\begin{itemize}
\item \textbf{(Term$\times$Term$\times$Article) Tensor (\TTA)}:
The most straightforward way that comes into mind for differentiating between a fake news and a real one is to analyze the content of the articles. Different classes of news articles, i.e., fake and real classes tend to have some common words that co-occur within the text. Thus, we use a tensor proposed by \cite{Hosseinimotlagh2017UnsupervisedCI} to model co-occurrence of these common words. 
This model not only represents content-based information but also considers the relations between the words and is stronger than widely used bag of words and tf/idf models. To this end, we first create a dictionary of all articles words and then slide a window across the text of each article and capture the co-occurred words. As a result, we will have a co-occurrence matrix for each article. By stacking all these matrices, we create a three mode tensor where the first two dimensions correspond to the indices of the words in the dictionary and the third mode indicates the article's ID. We may assign binary values or frequency of co-occurrence to entries of the tensors. However, as shown in \cite{ASONAM2018} binary tensor is able to capture more nuance patterns. So, in this work we also use binary values. 
\item \textbf{(Hashtag$\times$Term$\times$Article) Tensor (\HTA)}: Hashtags often show some trending across social media. Since, hashtags assigned to an article usually, convey social context information which is related to content of news article, we propose to construct a hashtag-content tensor to model such patterns. In this tensor, we want to capture co-occurrence of words within the articles and the hashtags assigned to them. For example, an article tagged with a hashtag \texttt{\#USElection2016} probably consists of terms like ``Donald Trump'' or ``Hillary Clinton''. These kind of co-occurrences are meaningful and convey some patterns which may be shared between different categories of articles. The first two modes of this tensor correspond to hashtag and word indices respectively and the third mode is article mode. %The entry values are also the binary co-occurrence of hashtags and words.
\item \textbf{(Article$\times$HTML features) Matrix (\Tags)}: Another source of information is the trustworthiness of serving webpage. In contrast to reliable web resources which usually have a standard form, misleading web pages may often be messy and full of different advertisement, pop-ups and multimedia features such as images and videos. Therefore, we suggest to create another model to capture the look and the feel of the web page serving the content of news article. We can approximate look of a web page by counting HTML features and tags and then represent it by a (article, HTML feature) matrix. The rows and columns of this matrix indicate the article and hashtag IDs respectively. We fill out the entries by frequency of HTML tags in the web source of each article domain. Fig. \ref{fig:jive} demonstrates aforementioned aspects.
\end{itemize}
\subsection{Hierarchical Decomposition}
Now, the goal is to find manifold patterns with respect to all introduced aspects. To this end, we look for shared components of the latent patterns.

\subsubsection{Level-1 decomposition: Finding article patterns with respect to each aspect}
 As mentioned above, first, we find the latent patterns of the articles with respect to each aspect. To do so, we decompose first two tensor models using Canonical Polyadic or CP/PARAFAC decomposition \cite{harshman1970fpp} into summation of rank one tensors, representing latent patterns within the tensors as follows: 
\begin{dmath}
{\mathcal{X}} \approx \Sigma_{r=1}^{R} \mathbf{a}_r \circ \mathbf{b}_r \circ \mathbf{c}_r
\end{dmath}
where $\mathbf{a}_r \in \mathbb{R}^{I}$, $\mathbf{b}_r \in \mathbb{R}^{J}$, $\mathbf{c}_r \in \mathbb{R}^{K}$
\cite{Papalexakis:2016} and ${R}$ is the rank of decomposition. The factor matrices are defined as $\mathbf{A} = [\mathbf{a}_1 ~ \mathbf{a}_2 \ldots \mathbf{a}_R]$, $\mathbf{B} = [\mathbf{b}_1 ~ \mathbf{b}_2 \ldots \mathbf{b}_R]$, and $\mathbf{C} = [\mathbf{c}_1 ~ \mathbf{c}_2 \ldots \mathbf{c}_R]$ where for \TTA and \HTA, $\mathbf{C}$ corresponds to the article mode and comprises the latent patterns of the articles with respect to content and social context respectively. The optimization problem for finding factor matrices is:
\begin{dmath}
 \min_{\mathbf{A,B,C}} ={{\lVert} \mathcal{X} - \Sigma_{r=1}^{R} \mathbf{a}_r \circ \mathbf{b}_r \circ \mathbf{c}_r{\rVert}}^2
\end{dmath}
 To solve the above optimization problem, we use Alternating Least Squares (ALS) approach which solves for any of the factor matrices by fixing the others \cite{Papalexakis:2016}. For the tag matrix, to keep the consistency of the model we suggest to use SVD decomposition ($
\mathbf{X} \approx\mathbf{U} \boldsymbol{\Sigma} \mathbf{V}^{T}$) because CP/PARAFAC is one extension of SVD for higher mode arrays. In this case, $\mathbf{U}$ comprises the latent patterns of articles with respect to overall look of the serving webpage.

\subsubsection{Level-2 decomposition: Finding manifold patterns}

Now, we want to put together article mode factor matrices resulted from individual decompositions to find manifold patterns with respect to all aspects. Let's $\mathbf{C_{TTA}}\in \mathbb{R} ^{N\times r_{1}}$ and  $\mathbf{C_{HTA}}\in \mathbb{R} ^{N\times r_{2}}$ be the third factor matrices resulted from CP decomposition of \TTA and \HTA respectively and let's $\mathbf{C_{TAGS}}\in \mathbb{R} ^{N\times r_{3}}$ be $\mathbf{U}$ matrix resulted from rank $r_{3}$ SVD of \Tags where $N$ is the number of articles. Since all these three matrices comprise latent patterns of the articles, we aim at finding patterns which are shared between all. To do so, we concatenate the three article embedding matrices ($\mathbf{C}$) and decompose the joint matrix of size $(r_1+r_2+r_3)\times N$ to find shared components. Like level-1 decomposition, we can simply take the SVD of joint matrix. SVD seeks an accurate representation in least-square setting but to find a meaningful representation we need to consider additional information  i.e., the relations between the components. Independent components analysis (ICA) tries to find projections that are statistically independent as follows:
\begin{dmath}
[\mathbf{C_{TTA}} ;\mathbf{C_{HTA}} ;\mathbf{C_{TAGS}}]=\mathbf{A}\mathbf{S} 
\end{dmath}
Where $\mathbf{A}$ is the corresponding shared factor of size $(r_1+r_2+r_3)\times N$ and $\mathbf{S}$ is the mixed signal of size $(r_1+r_2+r_3) \times R$. 
To achieve the statistical  independence, it looks for projections that leads to projected data to be as far from Gaussian distribution as possible. The problem is that, SVD (PCA) and ICA may result in identical subspaces. In fact, this happens when the direction of greatest variation and the independent components span the same subspace \cite{Vasilescu_thesis}

\par In order to consider relations between components and find a meaningful representation, we can consider shared and unshared components for $\mathbf{C}$ matrices such that unshared components are orthogonal to shared ones. To this end, we propose to use the Joint and Individual Variation Explained (JIVE) for level-2 decomposition\cite{JIVE}. More precisely, let's consider $\mathbf{A_i}$ and $\mathbf{J_i}$ to be the matrices representing the individual structure and submatrix of the joint pattern for $i\in\{ \HTA,\TTA,\Tags\}$ respectively such that they satisfy the orthogonality constraint. Using JIVE method, we decompose each article mode factor matrix $\mathbf{C}$ as follows:
\begin{dmath}
\mathbf{C_{i}}=\mathbf{J_{i}}+\mathbf{A_{i}}+\epsilon_{i}
\end{dmath}
To find $\mathbf{A}$ and $\mathbf{J}$ matrices, we use the approach presented in \cite{JIVE}. In other words, we fix $\mathbf{A}$ and find $\mathbf{J}$ that minimizes the following residual matrix:
\begin{dmath}
\norm{\mathbf{R}} ^2= \norm{\epsilon_{TTA} ;\epsilon_{HTA} ;\epsilon_{Tags}}^2
\end{dmath}

\begin{figure}[tb]
    \centering
    \includegraphics[width = 0.6\textwidth]{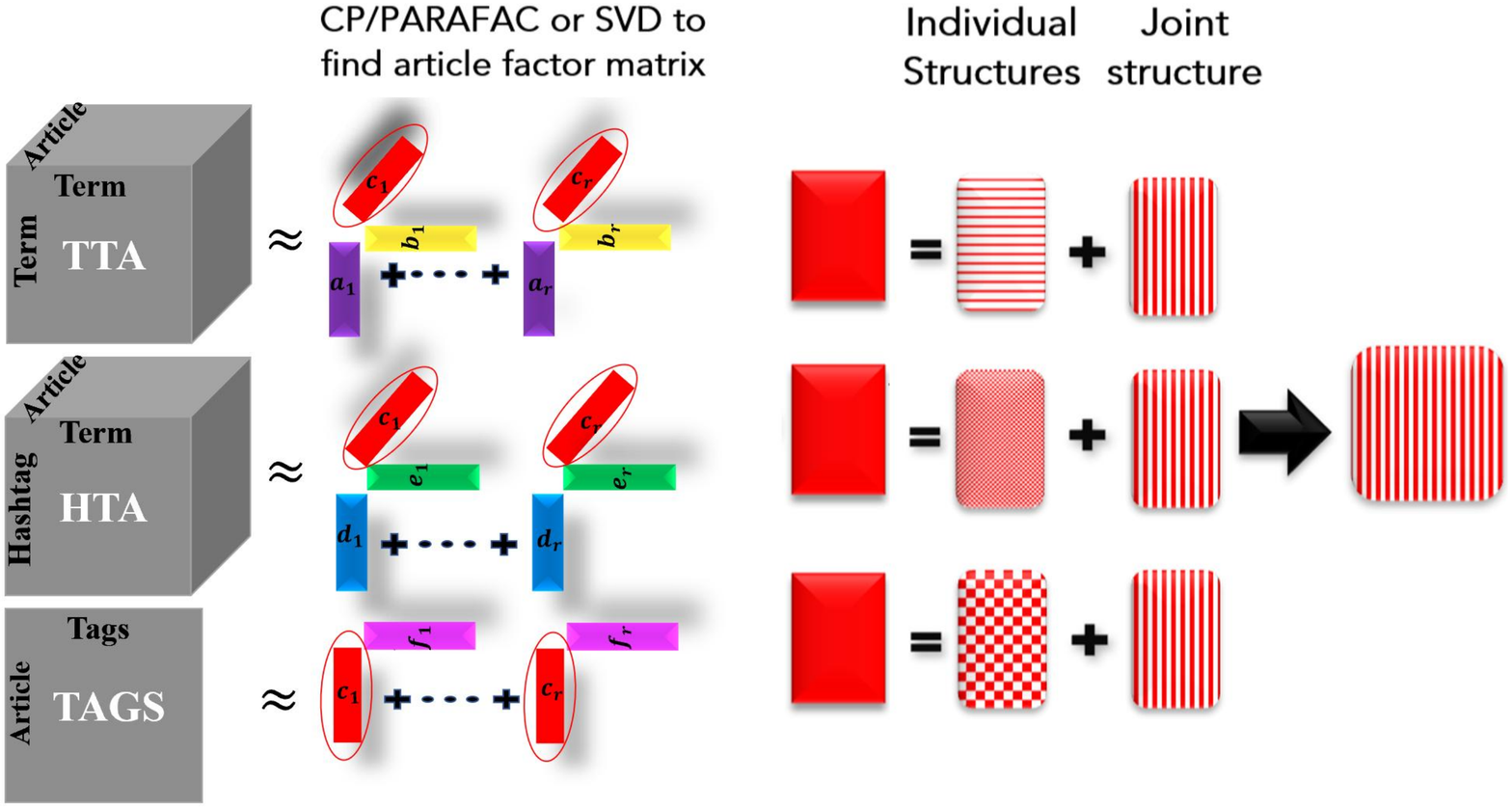}
    \caption{\CPJIVE finds manifold patterns of the articles that can be used for classification.} 
    \label{fig:jive}
\end{figure}

The joint structure $\mathbf{J}$ which minimizes $\norm{\mathbf{R}}^2$ is equal to the rank r SVD of joint matrix when we remove the individual structure and in the same way individual structure for each $\mathbf{C}$ matrix is the rank $r_i$ SVD of $\mathbf{C}$ matrix when we remove the joint structure \cite{JIVE}. Fig. \ref{fig:jive} and Algorithm. \ref{alg:1} demonstrate the details.

\begin{algorithm}[tb!]
\SetAlgoLined
\scriptsize
\algorithmicrequire{\TTA , \HTA , \Tags}
 \\\textbf{Output: $\mathbf{J_{joint}}$}
 \\\textbf{First level of Decomposition}
  \\$\mathbf{C_{TTA}}=$CP-ALS (\TTA,$r_1$);
  \\$\mathbf{C_{HTA}}=$CP-ALS (\HTA,$r_2$);
  \\$\mathbf{C_{Tags}}=$SVD (\Tags,$r_3$);
  \\$\mathbf{C^{Joint}}=[\mathbf{C_{TTA}} ;\mathbf{C_{HTA}} ;\mathbf{C_{Tags}}]$
\\\textbf{Second level of Decomposition}
\\ \While{$\norm{\mathbf{R}}^2 <\epsilon$}{
 {$\mathbf{J}=\mathbf{U}\mathbf{\Sigma} \mathbf{V}^T$}
\\ $[\mathbf{J_{TTA}} ;\mathbf{J_{HTA}} ;\mathbf{J_{Tags}}]=$SVD ($\mathbf{C_{Joint}},r_{joint}$)//Calculate $r_{joint}$ using Algorithm 2 
 \\ \For{$i\in\{$ \HTA,\TTA,\Tags$\}$}{
   $\mathbf{A_{i}}= \mathbf{C_{i}}-\mathbf{J_{i}}$
  \\ $\mathbf{A_{i}}=$SVD $(\mathbf{A_{i}}*(\mathbf{I}-\mathbf{V}\mathbf{V}^{T}),r_{i})$ //To satisfy the orthogonality constraint
   }
   $\mathbf{C_{i}}^{new}= \mathbf{C_{i}}^{joint}-\mathbf{A_{i}}$\\$\epsilon_i=|\mathbf{C_i}^{joint}-\mathbf{C_i}^{new}|$ 
 \\ $\mathbf{C}^{joint}=\mathbf{C}^{new}$
    }
\caption{\label{alg:1} \CPJIVE Hierarchical decomposition} 

\end{algorithm}

\begin{algorithm}[tb!]
\SetAlgoLined
\scriptsize
\algorithmicrequire{$\alpha \in (0, 1)$ , n\_perm}
\\\textbf{Output:} r
\\Let's $\lambda_{j}$ be the $j$’th singular value of $\mathbf{X} , i = 1,\dots, rank(\mathbf{X})$.
\\ \While{$n \leq n\_perm$}{
Permute the columns within each $X_i$, and calculate the singular values of the resulting  $\mathbf{C_{joint}}$
}
$\lambda_{i}^{perm}=100(1 - \alpha)$ percentile of $j$’th singular values
\\Choose largest $r$ such that $\forall {j} \leq r, \lambda_j \geq \lambda_j^{perm}$
\caption{\label{alg:2} Calculating the rank for joint and individual matrices}

\end{algorithm}

\subsection{Semi-Supervised Article Inference}
Previous step, provides us with a $N\times r$ matrix which comprises manifold patterns of $N$ articles. In this step, we leverage this matrix to address the semi-supervised problem of classifying misinformation. Row $i$ of this matrix represents article $i$ in $r$ dimensional space, we suggest to construct a K-NN graph using this matrix such that each node represents an article and edges are Euclidean distances between articles (rows of manifold patterns matrix) to model the similarity of articles.
We utilize the Fast Belief Propagation (FaBP) algorithm as is described in  \cite{ASONAM2018,KoutraKKCPF11} to propagate labels of known articles (fake or real) throughout K-NN graph. Belief propagation is a message passing-based algorithm. Let's ${m_{{j}\hookrightarrow{i}}}(x_i)$ indicates the message
 passes from node $j$ to node $i$ and ${N_i}$ denotes all the neighboring nodes of node $i$.
 ${m_{{j}\hookrightarrow{i}}}(x_i)$ conveys the opinion of node $j$ about the belief (label) of node $i$. Each node of a given graph $G$ leverages messages received from neighboring nodes and calculates its belief iteratively as follows:.
 
 \begin{dmath}
b_i(x_i)\propto \prod_{j\in    {(N_i)}}{m_{{j}\hookrightarrow{i}}}(x_i) 
\end{dmath}

\section{Experimental Evaluation}
In this section, we discuss the datasets, baselines and experimental results.
\subsection{Dataset Description}
\noindent{\bf \ourdata dataset}
To evaluate \CPJIVE, we created a new dataset by crawling Twitter posts contained links to articles and shared between June and August 2017. This dataset comprises 174k articles from more than 652 domains. For labeling the articles, we used BS-Detector, which is a {\em crowd-sourced} toolbox in form of a browser extension, as ground truth. BS-Detector categorizes domains into different categories such as bias, clickbait, conspiracy, fake, hate, junk science, rumor, satire, and unreliable. We consider above categories as ``misinformative'' class. A key caveat behind BS-Detector, albeit being the most scalable and publicly accessible means of labeling articles at-large, is that labels actually pertains to the {\em domain} rather than the article itself. At the face of it, this sounds like the labels obtained are for an entirely different task, however, Helmstetter et al. \cite{helmstetter2018weakly} show that training for this ``weakly labeled'' task (using labels for the domains), and subsequently testing on labels pertaining to the articles, yields minimal loss in accuracy and labels still hold valuable information. Thus, we choose BS-Detector for ground truth. However, in order to make our experiments as fair as possible, in light of the above fact regarding the ground truth, we do as follows:
\begin{itemize}
    \item We restrict the number of articles per domain we sample into our pool of articles. So, we experimented with randomly selecting a single article per domain and iterating over 100 such sets. Since we have 652 different domains in \ourdata dataset, in each iteration we chose 652 non-overlapping articles so, totally we examine different approaches for 65.2K different articles. 
  
 \item In order to observe the effect of using more articles per domain, we repeated the experiments for different number of articles per domain. We observed that the embedding that uses HTML tags receives a disproportionate boost in its performance. We attribute this phenomenon to the fact that all instances coming from the same domain have exactly the same HTML features, thus classifying correctly one of them implies correct classification for the rest, which is proportional to the number of articles per domain.
\item We balance the dataset so that we have 50\% fake and 50\% real articles at any given run per method. We do so to have a fair evaluation setting and prevent the situation in which there is a class bias. To show the insensitivity of \CPJIVE to class imbalance, we also experiment on an imbalanced dataset. 
\end{itemize}

\ \\
\noindent{\bf \politifact dataset}: For second dataset, we leverage FakeNewsNet dataset that the authors of \cite{Beyond} used for their experiments\footnote{\scriptsize https://github.com/KaiDMML/FakeNewsNet}\cite{shu2017fake,shu2017exploiting1}. This dataset consists of 1056 news articles from the \politifact fact checking website, 60\% being real and 40\% being fake. Using this imbalanced dataset, we can experiment how working on an imbalanced dataset may affect the proposed approach. Since not all of the signals we used from \ourdata dataset exist in \politifact dataset, For this dataset we created the following embeddings:

\begin{itemize}
\item \TTA tensor: To keep the consistency, we created \TTA from articles text. 
    \item User-News Interaction Embedding: We create a matrix which represent the users who tweets a specific news article, as proposed in \cite{Beyond}.
    \item Publisher-News Interaction Embedding: We create another matrix to show which publisher published a specific news article, as proposed in \cite{Beyond}.
\end{itemize}
Using the aforementioned signals, we can also test the efficacy of \CPJIVE when we leverage aspects other than those we proposed in this work.
\par \subsection{Baselines for Comparison}
As mentioned earlier, the two major contributions of \CPJIVE are: 1) the introduction of different aspects of an article and how they influence our ability to identify misinformation more accurately, and 2) how we leverage different aspects to find manifold patterns which belongs to different classes of articles. Thus, we conduct experiments with two categories of baseline to test each contribution separately: 
\\
\par \noindent{\bf Content-based approaches to test the effect of additional aspects.}
We compare with state-of-the-art content-based approaches to measure the effect of introducing additional aspects (hashtags and HTML features) into the mix, and whether the classification performance improves. We compare against:
\begin{itemize}
    \item \textbf{{\TTABP}}
    In \cite{ASONAM2018}, Bastidas et al. effectively use the co-occurrence tensor in a semi-supervised setting. They demonstrate how this tensor embedding outperforms other purely content-based state-of-the-art methods such as SVM on content-based features and Logistic regression on linguistic features \cite{hardalov_koychev_nakov_2016,Horne:2017}. Therefore, we select this method as the first baseline, henceforth referenced as ``\TTA''.
    The differences between our results and the results reported by Bastidas et al \cite{ASONAM2018} is due to using different datasets. However, since we used the publicly available code by Bastidas et al. \footnote{\scriptsize{https://github.com/Saraabdali/Fake-News-Detection-_ASONAM-2018}} \cite{ASONAM2018}, if we were to use the same data in \cite{ASONAM2018} the results would be exactly the same. 
    \item {\svm} is the well-known term frequency–inverse document frequency method widely used in text mining and information retrieval and illustrates how important a word is to a document. We create a \tfidf model out of articles text and apply SVM  classifier on the resulted model.
    \item \textbf{{\docvec}} is an NLP toolbox proposed by Le et al.\cite{doc2vec} from Google. This model is a shallow, two-layer neural network that is trained to reconstruct linguistic contexts of document. This algorithm is an extension to word2vec which can generate vectors for words. Since the SVM classifier is commonly used on this model, we also leverage SVM for document classification.\footnote{\scriptsize
   {https://github.com/seyedsaeidmasoumzadeh/Binary-Text-Classification-Doc2vec-SVM}}
    \item \textbf{{\fasttext}} is an NLP library by Facebook Research that can be used to learn word representations to efficiently classify document. It has been shown that \fasttext results are on par with deep learning models in terms of accuracy but an order of magnitude faster in terms of performance.\footnote{\scriptsize{https://github.com/facebookresearch/fastText}}
    \item \textbf{{\GloVeLSTM}} GloVe is an algorithm for obtaining vector representations of the words. Using an aggregated global word-word co-occurrence, this method results in a linear substructures of the word vector space. We use the method proposed in \cite{lin+al-2017-embed-iclr,lai2015recurrent} and we create a dictionary of unique words and leverage Glove to map indices of words into a pre-trained word embedding. Finally, as suggested, we use LSTM to classify articles.\footnote{\scriptsize{https://github.com/prakashpandey9/Text-Classification-Pytorch}}
\end{itemize} 

\begin{figure}[tb!]
\centering
    \includegraphics[width = 0.65\textwidth]{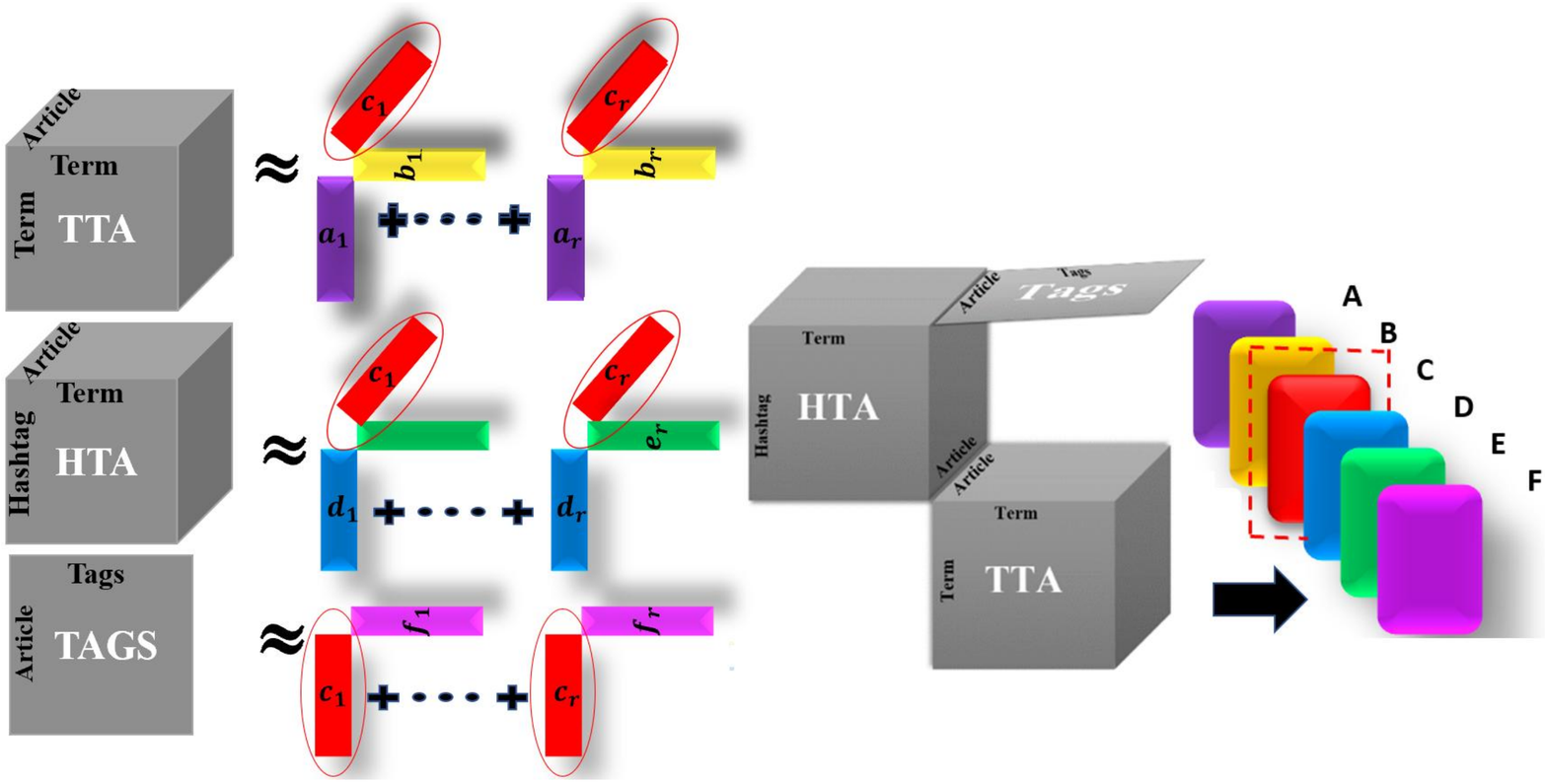}
    \caption{Couple Tensor Matrix Factorization for finding shared patterns of the articles.}
    \label{fig:CMTF}
\end{figure}

\vspace{1mm}
\noindent{\bf Ensemble approaches to test the efficacy of our fused method.}
Another way to jointly derive the article patterns, specially when the embeddings are tensors and matrices, is to couple matrices and tensors on shared mode(s) i.e., article mode in this work. In this technique which called coupled Matrix and Tensor Factorization (\CMTF), the goal is to find optimized factor matrices by considering all different optimization problems we have for individual embeddings. Using our proposed embeddings, we the optimization problem is:

\begin{dmath}
 \min_{\bf{A,B,C,D,E,F}} {{\lVert} {\TTA} -[\bf{A},\bf{B},\bf{C}]{\rVert}}^2+{\lVert} {\HTA}-[\bf{D},\bf{E}, \bf{C}]{\rVert}^2
  +{\lVert}{\Tags} -C{F}^T{\rVert}^2
  \label{equ:cmtf}
\end{dmath}

\noindent where $[\bf{A,B,C}]$ denotes $\Sigma_{r=1}^{R}\mathbf{a}_r\circ \mathbf{b}_r\circ \mathbf{c}_r$ , and $\bf{C}$ is the shared article mode as shown in Fig.\ref{fig:CMTF}. In order to solve the optimization problem above, we use the approach introduced in \cite{CMTF1}, which proposes {\em all-at-once-optimization} by computing the gradient of every variable of the problem, stacking the gradients into a long vector, and applying gradient descent. There is an advanced version of coupling called \ACMTF that uses weights for rank-one components to consider both shared and unshared ones\cite{ACMTF}. We applied \ACMTF as well and it led to similar results, however, slower than standard \CMTF. Thus, we just report the result of \CMTF.   

\par Moreover, as mentioned earlier, to derive the manifold pattern which is shared between the aspects, we can leverage different mathematical approaches such as: 
\begin{itemize}
    \item \SVD (Singular Value Decomposition) 
    \item \JICA (Joint Independent Component Analysis)
\end{itemize}
 on $\mathbf{C_{joint}}$. To measure the performance of \JIVE method for finding manifold patterns, we will also examine the above approaches for level-2 decomposition.

\subsection{Comparing with Baselines}

\subsubsection{Implementation.} We implemented \CPJIVE and all other approaches in MATLAB using Tensor Toolbox version 2.6. Moreover, to implement \JICA approach, we used FastICA\footnote{\scriptsize https://github.com/aludnam/MATLAB/tree/master/FastICA\_25} , a fast implementation of ICA. For the baseline approach \CMTF we used the Toolbox\footnote{\scriptsize http://www.models.life.ku.dk/joda/CMTF\_Toolbox} in \cite{CMTF1,CMTF2}. For the belief propagation step, we used implementation introduced in \cite{KoutraKKCPF11}. Based on the experiments reported in \cite{ASONAM2018}, we employed a sliding window of size 5 for capturing the co-occurring words. Using AutoTen \cite{papalexakis2016autoten}, which finds the best rank for tensors, we found out the best rank for \TTA and \HTA is 10 and 40 respectively. For the \Tags embedding, we took the full SVD to capture the significant singular values and set the \Tags SVD rank to 20. Since for \CMTF approach we have to choose the same rank, as a heuristic, we used the range of ranks for individual embeddings (10 for \HTA and 40 for \TTA) and again searched for the ensemble rank in this range. Based on our experiments, rank 30 leads to the best results in terms of F1-score. Moreover, grid search over 1-30 nearest neighbors yielded choice of 5 neighbors for \CMTF and 15 for \CPJIVE. For the rank of joint model as well as individual and joint structures i.e., $\mathbf{A_i}$ and $\mathbf{J_i}$, we used the strategy proposed in \cite{JIVE} and for reproducibility purposes is demonstrated in Algorithm \ref{alg:2}. 
The intuition is to find the rank of joint structure $i$ by comparing the singular values of the original matrix with the singular values of $n_{perm}$ randomly permuted matrices. If the $j^{th}$ singular value in the original matrix is $\geq 100 (1-\alpha)$ percentile of the $j^{th}$ singular
value of $n_{perm}$ matrices, we keep it as a significant one. Number of these significant singular values shows the rank. $n_{perm}$ and $\alpha$ are usually set to 100 and 0.05 respectively. For timing experiment, we used the following configuration:
\begin{itemize}
    \item CPU: Intel(R) Core(TM) i5-8600K CPU @3.60GHz
    \item OS: CentOS Linux 7 (Core)
    \item RAM: 40GB
\end{itemize}

\subsubsection{Testing the effect of different aspects.} 
This experiment refers to the first category of baselines i.e., state-of the-art content-based approaches introduced earlier. Table \ref{table:content-based} demonstrates the comparison; label\% shows the percentage of known data used for propagation/training of the models. As reported, \CPJIVE outperforms all content-based approaches significantly. For example, using only 10\% of known labels the F1 score of \CPJIVE is around 12\% and 13\% percent more than \docvec and \fasttext respectively. In case of \GloVeLSTM, due to small size of training set i.e., 10 and 20 percent, the model overfits easily which shows the strength of \CPJIVE against deep models when there is scarcity of labeled data for training. Moreover, our ensemble model that leverages \TTA as one of its embeddings beats the individual decomposition of \TTA which illustrates the effectiveness of adding other aspects of the data for modeling the news articles.
\iffalse
\begin{table*}[!h]
\centering
\scriptsize
\begin{tabular}{c c c c c c c}
 \toprule
 \%labels &\svm &\docvec &\fasttext & \GloVeLSTM &\TTA &\CPJIVE\\
 \toprule
 10 &% 0.3797$\pm$0.0037&
 0.5002$\pm$ 0.0326&0.5710$\pm$0.0929&0.5628$\pm$0.0310& -%0.4320$\pm$  0.1501
 &0.5821$\pm$0.0180&\textbf{0.6932}$\pm$ \textbf{0.0098}\\
 20 &%0.3895$\pm$0.0079&
 0.4617$\pm$0.0134&0.5586$\pm$0.0676&0.5735$\pm$0.0275 &-%0.4761$\pm$0.1255
 &0.5988$\pm$0.0188& \textbf{0.7171}$\pm$ \textbf{0.0105}\\
 30 &%0.4106$\pm$0.0146&
 0.4643$\pm$0.0158& 0.5482$\pm$0.0480&0.5860$\pm$0.0242 & 0.5024$\pm$0.0876&0.6093$\pm$0.0199& \textbf{0.7320}$\pm$ \textbf{0.0113}\\
 40 &%0.4716$\pm$0.0228&
 0.4756$\pm$0.0231&0.5470$\pm$0.0342& 0.5927$\pm$0.0285 & 0.5031$\pm$0.0609&0.6143$\pm$0.0223& \textbf{0.7408}$\pm$ \textbf{0.0106}\\
\bottomrule
\end{tabular}
\caption{ F1 score of \CPJIVE against state-of-the-art content-based methods on \ourdata dataset.}
\label{table:content-based}
\end{table*}
\fi
\begin{table*}[tb!]
\centering
\tiny
\addtolength{\tabcolsep}{2pt}
\begin{tabular}{c c c c c c c}
 \toprule
 \textbf{\scriptsize \%Labels}&\textbf{\scriptsize \svm}&\textbf{\scriptsize \docvec}&\textbf{\scriptsize \fasttext}&\textbf{\scriptsize \GloVeLSTM}&\textbf{\scriptsize \TTABP}&\textbf{\scriptsize \CPJIVE}\\
 \toprule
 % 0.3797$\pm$0.0037& 
 10&0.500$\pm$0.032&0.571$\pm$0.092&0.562$\pm$0.031& -%0.4320$\pm$  0.1501
 &0.582$\pm$0.018&\textbf{0.693}$\pm$\textbf{0.009}\\
 20 &%0.3895$\pm$0.0079&
 0.461$\pm$0.013&0.558$\pm$0.067&0.573$\pm$0.027&-%0.4761$\pm$0.1255
 &0.598$\pm$0.018&\textbf{0.717}$\pm$\textbf{0.010}\\
 30 &%0.4106$\pm$0.0146&
 0.464$\pm$0.015& 0.548$\pm$ 0.048&0.586$\pm$0.024& 0.502$\pm$0.087&0.609$\pm$0.019& \textbf{0.732}$\pm$\textbf{0.011}\\
 40&%0.4716$\pm$0.0228&
 0.475$\pm$0.023&0.547$\pm$0.034&0.592$\pm$ 0.028& 0.503$\pm$0.060&0.614$\pm$0.022& \textbf{0.740}$\pm$\textbf{0.010}\\
\bottomrule
\end{tabular}
\caption{F1 score of \CPJIVE outperforms all content-based methods on \ourdata dataset.}
\label{table:content-based}
\end{table*}
%%%%%%%%%%%%%%%%%%%%%%%%%%%%%%%%%%%%%%%

\subsubsection{Testing the efficacy of fused method in \CPJIVE.}
\begin{table*}[tb!]
\centering
\tiny
\addtolength{\tabcolsep}{2pt}
\begin{tabular}{ c c c c c c}
\toprule
 &\multicolumn{3}{c}{\textbf{\scriptsize \ourdata}} & \multicolumn{2}{c}{\scriptsize \textbf{\politifact}}\\
\cmidrule(lr){2-4}
\cmidrule(lr){5-6}
\textbf{\scriptsize \%Labels}& \textbf{\scriptsize \CMTF}&\textbf{\scriptsize \CMTFpp} %& \Stacking%
 &\textbf{\scriptsize \CPJIVE}&\textbf{\scriptsize \CMTF}&\textbf{\scriptsize \CPJIVE}\\
\midrule
10&0.657$\pm$0.009&0.657$\pm$0.010 %&0.6732$\pm$0.0075
&\textbf{0.693}$\pm$\textbf{0.009} &0.733$\pm$0.007%&0.7283&
&\textbf{0.766}$\pm$\textbf{0.007}\\
20&0.681$\pm$0.010&0.681$\pm$0.009 %&0.6963$\pm$0.0076
&\textbf{0.717}$\pm$ \textbf{0.010}&0.752$\pm$0.012%0.7562
&\textbf{0.791}$\pm$\textbf{0.007}\\
30&0.691$\pm$0.010&0.692$\pm$0.009 %&0.7079$\pm$0.0076
&\textbf{0.732}$\pm$\textbf{0.011}& 0.774$\pm$0.006%&0.7774&
&\textbf{0.802}$\pm$\textbf{0.007}\\
40&0.699$\pm$0.010&0.699$\pm$0.009 %&0.7145$\pm$0.0079
&\textbf{0.740}$\pm$\textbf{0.010}&0.776$\pm$0.004
&\textbf{0.810}$\pm$\textbf{0.008}\\
\bottomrule
\end{tabular}
\caption{\CPJIVE outperforms coupling approaches in terms of F1 score in both  datasets.}
\label{table:ourdata_cmtf}
\end{table*}

For the second category of baselines, we compare \CPJIVE against the recent work for joint decomposition of tensors/matrices i.e., \CMTF. As discussed earlier, we couple \TTA, \HTA and \Tags on shared article mode. Since the "term" mode is also shared between \TTA and \HTA, we also examine the \CMTF by coupling on both article and term modes, henceforth referenced as \CMTFpp in the experimental results. For the \politifact dataset, there is only one shared mode i.e., article mode. therefore, we only compare against \CMTF approach. The experimental results for theses approaches are shown in Table \ref{table:ourdata_cmtf}. As illustrated, \CPJIVE leads to higher F1 score which confirms the effectiveness of \CPJIVE for jointly classification of articles.  One major drawback of \CMTF approach is that, we have to use a unique decomposition rank for the joint model which may not fit all embeddings and  may lead to losing some informative components or adding useless noisy components due to inappropriate rank of decomposition. Another drawback of this technique is that, as we add more embeddings, the optimization problem becomes more and more complicated which may cause the problem become unsolvable and infeasible in terms of time and resources. The time efficiency of \CPJIVE against \CMTF approach is reported in Table. \ref{table:timing};  we refer to level-1 and level-2 decompositions as L1D and L2D respectively. As shown, for all ranks, \CPJIVE is an order of magnitude faster than \CMTF due to the simplicity of optimization problem in comparison to equation \ref{equ:cmtf} which means \CPJIVE is more applicable for real world problems. 

\begin{table}[tb!]
\centering
\tiny
\addtolength{\tabcolsep}{2pt}
\begin{tabular}{ c c c c c c c}
\toprule
 \multicolumn{1}{c}{} &
 \multicolumn{2}{c}{\textbf{\scriptsize L1D}} & \multicolumn{2}{c}{\textbf{\scriptsize L2D+Rank finding}} &\multicolumn{2}{c}{\textbf{\scriptsize Total time (Secs.)}} \\
 \cmidrule(r){2-3}
\cmidrule(r){4-5}
\cmidrule(r){6-7}
\centering
\textbf{\scriptsize L1D
Rank}&\textbf{\scriptsize \CMTF}&\textbf{\scriptsize \CP/\SVD} &\textbf{\scriptsize \CMTF}&\textbf{\scriptsize \JIVE}&\textbf{\scriptsize \CMTF}&\textbf{\scriptsize \CPJIVE}\\
\midrule
5&352.96&7.63&-& 13.69&352.96&21.32\\
10&1086.70&16.35&-&49.69 &1086.70&66.04\\
20&11283.51&44.25&-&133.70 &11283.51&177.95\\
30&13326.80&84.76&-&  278.43&13326.80&363.19\\

\bottomrule
\end{tabular}
\caption{Comparing execution time (Secs.) of \CMTF against \CPJIVE on \ourdata dataset shows that \CPJIVE is an order of magnitude faster than \CMTF approach.}
\label{table:timing}
\end{table}

\subsubsection{Testing the efficacy of using \JIVE for level-2 decomposition.}
In this experiment we want to test the efficacy of \JIVE against other approaches for deriving the joint structure of $\mathbf{C}$. The evaluation results for this experiment are reported in Table.  \ref{table:3approaches}. As shown, \SVD and \JICA resulted in same classification performance which as explained earlier indicates that the directions of greatest variation and the independent components span the same subspace \cite{Vasilescu_thesis}. However, the F1 score of \CPJIVE is higher than two other methods on both datasets which practically justifies that considering orthogonal shared and unshared parts and minimizing the residual can improve the performance of naive SVD.
\iffalse
\begin{table}[tb!]
\centering
\scriptsize
\begin{tabular}{c c c c c c c}
\toprule
\multicolumn{1}{c}{\%labels}& \multicolumn{3}{c}{\ourdata} & \multicolumn{3}{c}{\politifact}\\
\cmidrule(lr){1-1}
\cmidrule(lr){2-4}
\cmidrule(lr){5-7}
&\CPICA&\CPSVD&\CPJIVE&\CPICA&\CPSVD&\CPJIVE\\
\midrule
10&0.6845\pm0.0092 & 0.6859\pm0.0175 & \textbf{0.6932}\pm \textbf{0.0098}
&0.7490\pm0.0064&0.7568\pm0.0092&\textbf{0.7665}\pm \textbf{0.0076}\\
20&0.7102\pm0.0099 &0.7126\pm0.0145 & \textbf{0.7171} \pm \textbf{0.0105}
&0.7754\pm0.0065&0.7833\pm0.0099 &\textbf{0.7917}\pm\textbf{0.0073}\\
30&0.7248\pm0.0095 & 0.7245\pm0.0185 &\textbf{0.7320}\pm \textbf{0.0113}
&0.7863\pm0.0064&0.7963\pm0.0121&\textbf{0.8024}\pm \textbf{0.0077}\\
40& 0.7341\pm0.0096 & 0.7357\pm0.0121&\textbf{0.7408}\pm \textbf{0.0106}
&0.7971\pm0.0060&0.8070\pm0.0081&\textbf{0.8106}\pm\textbf{0.0088}\\
\bottomrule
\end{tabular}
\caption{ F1 score of proposed approaches on \ourdata and \politifact datasets.} 
\label{table:3approaches}
\end{table}
\fi
\begin{table}[tb!]
\centering
\tiny
\addtolength{\tabcolsep}{2pt}
\begin{tabular}{c c c c c c c}
\toprule
\multicolumn{1}{c}{\textbf{\scriptsize \%Labels}}& \multicolumn{3}{c}{\textbf{ \scriptsize \ourdata}} & \multicolumn{3}{c}{\textbf{ \scriptsize \politifact}}\\
\cmidrule(lr){1-1}
\cmidrule(lr){2-4}
\cmidrule(lr){5-7}
&\textbf{ \scriptsize \JICA}&\textbf{ \scriptsize \SVD}&\textbf{\scriptsize \JIVE}&\textbf{\scriptsize \JICA}&\textbf{\scriptsize \SVD}&\textbf{\scriptsize \JIVE}\\
\midrule
10&0.684$\pm$0.009&0.685$\pm$0.017& \textbf{0.693}$\pm$\textbf{0.009}
&0.749$\pm$0.006&0.756$\pm$0.009&\textbf{0.766}$\pm$\textbf{0.007}\\
20&0.710$\pm$0.009&0.712$\pm$0.014& \textbf{0.717}$\pm$\textbf{0.010}
&0.775$\pm$0.006&0.783$\pm$0.009 &\textbf{0.791}$\pm$\textbf{0.007}\\
30&0.724$\pm$0.009& 0.724$\pm$0.018 &\textbf{0.732}$\pm$\textbf{0.011}
&0.786$\pm$0.006&0.796$\pm$0.012&\textbf{0.802}$\pm$\textbf{0.007}\\
40& 0.734$\pm$0.009 & 0.735$\pm$0.012&\textbf{0.740}$\pm$\textbf{0.010}
&0.797$\pm$0.006&0.807$\pm$0.008&\textbf{0.810}$\pm$\textbf{0.008}\\
\bottomrule
\end{tabular}
\caption{Comparing the efficacy of different approaches for level-2 decomposition illustrates that using \JIVE approach for \CPJIVE ouperforms other methods.}

\label{table:3approaches}
\end{table}
\subsubsection{\CPJIVE vs. single aspect modeling.}
\begin{figure}[!h]
\centering
\includegraphics[width = 0.46\textwidth]{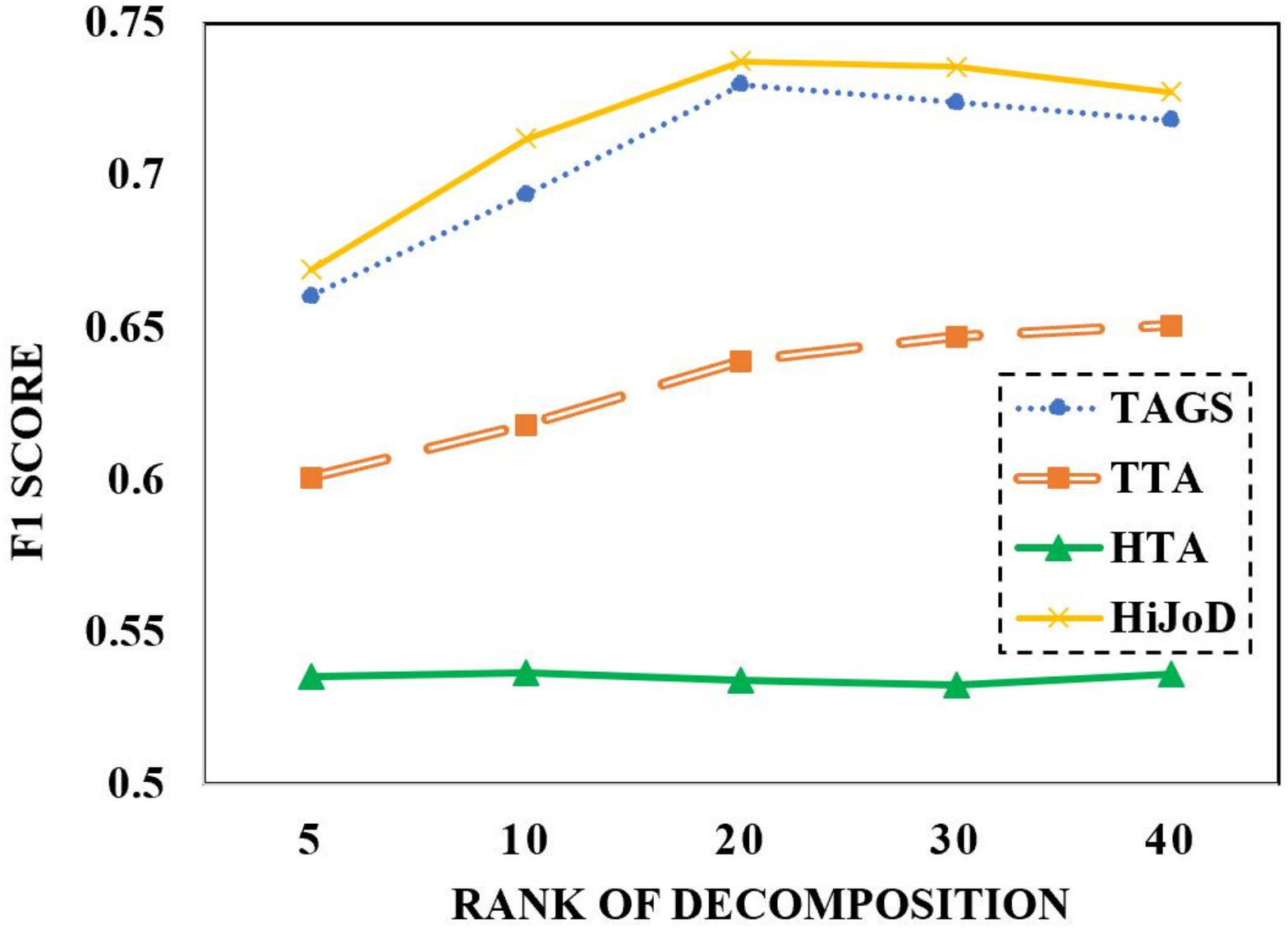}
\caption{F1-score of using individual embeddings vs. \CPJIVE. Even when the best rank of each embedding is not used, \CPJIVE outperforms individual decompositions} 
\label{fig:individual}
\end{figure}

Finally, we want to investigate how adding other aspects of the articles affect the classification performance. To this end, we decompose our proposed embeddings i.e., \TTA, \HTA and \Tags extracted from \ourdata dataset 
individually and leverage the $\mathbf{C}$ matrices for classification. The result of this experiment is demonstrated in Fig. \ref{fig:individual}. As mentioned before, in contrast to \CMTF, in \CPJIVE we can use different ranks for different aspects as we did so previously. However, in order to conduct a fair comparison between individual embeddings and \CPJIVE, we merge the embeddings of the same rank. It is worth mentioning that, performance of \CPJIVE is higher than what is shown in Fig. \ref{fig:individual} due to concatenation of $\mathbf{C}$ matrices of the best rank. So, this result is just for showing the effect of merging different aspects by fixing other parameters i.e., rank and $k$. As shown, the \CPJIVE even when we join embeddings of the same rank and do not use the best of which outperforms individual decompositions.

\section{Related Work}
\noindent{\bf Ensemble modeling for misinformation detection.}
A large number of misinformation detection approaches focus on a single aspect of the data, such as article content \cite{wu2017gleaning,ASONAM2018}, user features \cite{wu2018tracing}, and temporal properties \cite{kumar2018false}. There exist, however, recent approaches that integrate various aspects of an article in the same model. For example, in \cite{Beyond} the authors propose an ensemble model for finding fake news. In this approach, a bag of words embedding is used to model content-based information, while in this work, we leverage a tensor model i.e., \TTA which not only enables us to model textual information, but also is able to capture nuanced relations between the words. The different sources of information used in \cite{Beyond} (user-user, user-article and publisher-article interactions) do not overlap with the aspects introduced here (hashtags and HTML features), however, in our experiments we show that \CPJIVE effectively combines both introduced aspects as well as the ones in \cite{Beyond}. In another work \cite{dEFEND}, news contents and user comments are exploited jointly to detect fake news. Although user comment is a promising aspect, still the main focus is on the words of comments. However, we use HTML tags and hashtags in addition to the textual content. 
   \\{\bf Semi-supervised Learning / Label Propagation Models.}
    The majority of mono-aspect modeling proposed so far leverage a supervised classifier. For instance, in  \cite{hardalov_koychev_nakov_2016}
    a logistic regression classifier is used which employs linguistic and semantic features for classification. In \cite{Horne:2017}, authors apply a SVM classifier for content based features. Moreover, some works have been done using recurrent neural network (RNN) and Dynamic Series-Time Structure (DSTS) models \cite{DBLP:journals/corr/RuchanskySL17,Ma:2015}.  In contrary to the aforementioned works, we use a model which achieves very precise classification when leverage very small amount of ground truth. 
    There are some proposed methods that mainly rely on propagation models. For example, in \cite{Jin:2014} the authors proposed a hierarchical propagation model on a suggested three-layer credibility network. In this work, a hierarchical structure is constructed using event, sub-event and message layers, even though a supervised classifier is required to obtain initial credibility values. In \cite{NewsVerif}, a PageRank-like credibility propagation method is proposed to apply on a network of events, tweets and users. In this work, we leverage belief propagation to address the semi-supervised problem of misinformation detection. We show that proposed approach outperforms state-of-the-art approaches in label scarcity settings. 
    
\label{sec:related}
\section*{Acknowledgments}
The authors would like to thank Gisel Bastidas for her invaluable help with data collection. Research was supported by a UCR Regents Faculty Fellowship, a gift from Snap Inc., the Department of the Navy, Naval Engineering Education Consortium under award no. N00174-17-1-0005, and the National Science Foundation Grant no. 1901379. Any opinions, findings, and conclusions or recommendations expressed in this material are those of the author(s) and do not necessarily reflect the views of the funding parties.

\section{Conclusion}
\label{sec:conclusions}
In this paper, we propose \CPJIVE, a 2-level decomposition pipeline that integrate different aspects of an article towards more precise discovery of misinformation on the web. Our contribution is two-fold: we introduce novel aspects of articles which we demonstrate to be very effective in classifying misinformative vs. real articles, and we propose a principled way of fusing those aspects leveraging tensor methods.
We show that \CPJIVE is not only able to detect misinformation in a semi-supervised setting even when we use only 10\% of the labels but also an order of magnitude faster than  similar ensemble approaches in terms of execution time. Experimental results illustrates that \CPJIVE achieves F1 score of roughly 74\% and 81\% on \ourdata and \politifact datasets respectively which outperforms state-of-the-art content-based and neural network based approaches. 
\balance

\end{document}